\documentstyle[11pt]{article}

\def\theequation{\arabic{section}.\arabic{equation}}

\renewcommand{\theequation}{\thesection.\arabic{equation}}

\global\arraycolsep=1pt 
\input epsf
\input{tcilatex}
\begin{document}

\font\cmss=cmss10 \font\cmsss=cmss10 at 7pt \hfill \hfill IFUP-TH/01-01


\vspace{10pt}

\begin{center}
{\Large {\bf \vspace{10pt} A UNIVERSAL FLOW INVARIANT\\[0pt]
\vspace{5pt} IN QUANTUM FIELD THEORY }} \vspace{10pt}

\bigskip \bigskip

{\sl Damiano Anselmi}

{\it Dipartimento di Fisica, Universit\`a di Pisa, via F. Buonarroti 2,
56126 Pisa, Italia}
\end{center}

\vskip 2truecm

\begin{center}
{\bf Abstract}
\end{center}

A flow invariant is a quantity depending only on the UV and IR conformal
fixed points and not on the flow connecting them. Typically, its value is
related to the central charges $a$ and $c$. In classically-conformal field
theories scale invariance is broken by quantum effects and the flow invariant 
$a_{{\rm UV}}-a_{{\rm IR}}$ is measured by the area of the graph of the beta
function between the fixed points. There exists a theoretical explanation of
this non-trivial fact. On the other hand, when scale invariance is broken at
the classical level, it is empirically known that the flow invariant equals $%
c_{{\rm UV}}-c_{{\rm IR}}$ in massive free-field theories, but a theoretical
argument explaining why it is so is still missing. A number of related open
questions are answered here. A general formula of the flow invariant is
found, which holds also when the stress tensor has improvement terms. The
conditions under which the flow invariant equals $c_{{\rm UV}}-c_{{\rm IR}}$
are identified. Several non-unitary theories are used as a laboratory, but
the conclusions are general and an application to the Standard Model is
addressed. The analysis of the results suggests some new minimum principles,
which might point towards a new understanding of quantum field theory.

\vspace{4pt}

\vskip 1truecm

{Pacs: 11.25.H; 11.10.Gh; 11.15.Bt; 11.40.Ex; 04.62.+v} \vfill\eject

\section{Introduction}

\setcounter{equation}{0}

Anomalies are among the most powerful tools for the investigation of quantum
field theory beyond the perturbative expansion. The Adler-Bardeen theorem 
\cite{adler} and the 't Hooft anomaly matching conditions \cite{thooft} give
exact information about the strongly interacting limit of the theory. The
trace anomaly encodes the beta function and therefore the
renormalization-group flow. Moreover, various anomalies are computable to
high orders in perturbation theory, with a limited effort, and in various
models, combining the Adler-Bardeen theorem with supersymmetry, they are
computable exactly to all orders \cite{noi,noi2}. All-order formulas are
available also in non-supersymmetric theories, for example the formula for
the RG flow of the anomaly called $a$ \cite{athm,at6d}.

The quantity $a$ is one of the two anomaly coefficients of the trace $\Theta 
$ of the stress tensor in external gravity. It multiplies the Euler density.
The other anomaly, called $c$, is the coefficient of the square of the Weyl
tensor. (This terminology was introduced in \cite{noi,central}). The
quantity $c$ is also the coefficient of the stress-tensor two-point
function. The definition of $a$ and $c$ by means of the trace anomaly in
external gravity is meaningful only in even dimensions. In odd dimensions,
the trace anomaly in external gravity is not useful, but $c$ can still be
defined by means of the stress-tensor two-point function.

In two dimensions, the difference $c_{{\rm UV}}-c_{{\rm IR}}$ between the
critical values of the unique central charge is related to the correlator $%
\langle \Theta \,\Theta \rangle $ \cite{zamolo,cardy}. Precisely, the
formula reads 
\begin{equation}
c_{{\rm UV}}-c_{{\rm IR}}=3\pi\int {\rm d}^{2}x\,|x|^{2}\,\langle \Theta
(x)\,\Theta (0)\rangle .  \label{fl2d}
\end{equation}
This expression is an example of flow invariant, that is to say a quantity
defined as the integral of a correlator along the flow\footnote{
In (\ref{fl2d}) and in the other flow integrals appearing in 
the paper, the integrand 
is taken at distinct points. One can exclude a circle of radius
$\epsilon$ centered in $x=0$ and take the limit $\epsilon\rightarrow 0$
after the integration.}, whose
value depends only on the end points of the flow. The purpose of this paper
is to
investigate the existence of a universal flow invariant in four dimensions.

The two-dimensional property does not generalize immediately to higher
dimensions, where the issues are more involved. A priori, the
correlator $\langle \Theta \,\Theta \rangle $ should not know about either $%
a $ or $c$, in dimension greater than two. In reality, it knows
about both. For example, in marginally relevant flows, generated by the
dynamical scale $\mu $ in classically conformal quantum field
theories (strictly renormalizable at the quantum level), a theoretical
argument \cite{athm}, based on a physical principle, shows that the
integral $\int {\rm d}^{n}x\,|x|^{n}\,\langle \Theta (x)\,\Theta (0)\rangle $
is proportional to $a_{{\rm UV}}-a_{{\rm IR}}$. On the other hand, empirical
evidence suggests that the same integral is relevant also when conformality
is broken at the classical level, such as in the presence of masses or
super-renormalizable parameters (relevant flows). The integral, however, is
proportional to $c_{{\rm UV}}-c_{{\rm IR}}$ in massive free-field theories 
\cite{cappelli?}. These facts suggest that the integral $\int {\rm d}%
^{n}x\,|x|^{n}\,\langle \Theta (x)\,\Theta (0)\rangle $ is the basic
ingredient for the constrution of the universal flow invariant in four
dimensions, but a number of puzzles are raised, which are considered in this
paper.

In the study of flow-invariants in four dimensions, marginally
relevant and relevant flows exhibit crucially different properties. In the
case of a marginally relevant flow, the trace of the stress tensor is an
evanescent operator; in a more general relevant flow it is not. For example,
in Yang-Mills theory or massless QCD, we have 
\[
\Theta ={\frac{1}{4}}\varepsilon F_{{\rm B}}^{2}={\frac{1}{4}}\beta F_{{\rm R%
}}^{2}, 
\]
where $\beta =\partial \ln \alpha /\partial \ln \mu $, $\varepsilon =4-n$
and $F_{{\rm B}}^{2}$, $F_{{\rm R}}^{2}$ denote the bare and renormalized
operators, respectively. Classical conformal invariance means that $\Theta $
is generated only by quantum effects. In the presence of
super-rinormalizable interactions or masses, we have additional terms of the
form 
\[
\Theta =-m^{2}\phi ^{2}-m\bar{\psi}\psi , 
\]
and $\Theta $ is nonzero at the classical level.

This might not seem an important difference, at first sight. However, the
arguments of \cite{athm} depend crucially on the evanescence of the operator 
$\Theta $. This very evanescence guarantees that the induced action for the
conformal factor is convergent. To introduce the research of this paper, it
is compulsory to recall those arguments in some detail. Then I list the open
questions and describe the answers found in this paper. I also give some
examples of physical applications.

\subsection{The intrinsic difference between marginally relevant 
and strictly relevant flows}

I consider the four-dimensional case, for simplicity. To study correlators
of the stress tensor, the theory is embedded in an external gravitational
field. Here we are interested in the dependence on the conformal factor $%
\phi $ of the metric $g_{\mu\nu}$ and set $g_{\mu \nu }={\rm e}^{2\phi
}\delta _{\mu \nu }$. The induced action for the conformal factor reads at
the critical points 
\begin{equation}
S_{{\rm E}}[\phi ]=\frac{1}{180}\frac{1}{(4\pi )^{2}}\int {\rm d}%
^{4}x\left\{ a_{*}(\Box \phi )^{2}-(a_{*}-a_{*}^{\prime })\left[ \Box \phi
+(\partial _{\mu }\phi )^{2}\right] ^{2}\right\} .  \label{lor}
\end{equation}
where $a_{*}$ and $a_{*}^{\prime }$ are defined by the trace anomaly: 
\[
\Theta =\frac{1}{90(4\pi )^{2}}\left[ a_{*}{\rm e}^{-4\phi }\Box ^{2}\phi +%
\frac{1}{6}(a_{*}-a_{*}^{\prime })\Box R\right] . 
\]

Off-criticality, the induced action has additional non-local contributions.
In particular, we consider the $\Theta $-two-point function $\langle \Theta
(x)\,\Theta (0)\rangle $. The off-critical structure of the correlator 
depends on whether $\Theta $ is evanescent or not.

In a classically-conformal quantum field theory, where $\Theta $ is
evanescent, the correlator has the form \cite{athm} 
\begin{equation}
\langle \Theta (x)\ \Theta (0)\rangle =\frac{1}{180\pi ^{2}}\frac{1}{(4\pi
)^{2}}\Box ^{2}\left( \frac{\beta ^{2}(t)\tilde{f}(t)}{|x|^{4}}\right) ,
\label{correction}
\end{equation}
where $t=\ln |x|\mu $. This correlator is convergent for $|x|\rightarrow 0$.
The perturbative expansion of (\ref{correction}) has the form 
\begin{equation}
\langle \Theta (x)\ \Theta (0)\rangle =\Box ^{2}\sum_{k}a_{k}(g)\left( \frac{%
t^{k}}{|x|^{4}}\right) ,
\end{equation}
where $g$ is the coupling constant. The coefficients $a_{k}(g)$ are of order 
$g^{2(n+1)} $. The poles for $|x|\rightarrow 0$ are infinitely many,
classified by the powers $t^{k}/|x|^{4}$. For example, $1/|x|^{4}$ has a
simple pole in the limit $|x|\rightarrow 0$, $\ln( |x|\mu) /|x|^{4}$ has a
double and a simple pole, and so on. However, the poles resum together into
the beta function, which carries an additional zero for $|x|\rightarrow 0$.
It can be shown that this zero produces the desired convergence \cite{athm}.

On the other hand, if $\Theta $ is not evanescent, such as in the case of a
massive free scalar field $\varphi $, then the correlator has the form 
\[
\langle \Theta (x)\ \Theta (0)\rangle ={\frac{m^{6}}{16\pi ^{4}|x|^{2}}}%
K_{1}^{2}(m|x|), 
\]
where $K_{1}$ is the modified Bessel function. There is just one pole, $%
1/(8\pi ^{4})\,\,m^{4}/|x|^{4},$ for $|x|\rightarrow 0$. Therefore, no
cancellation can occur. We see that in the class of problems we are
considering, classically-conformal quantum field theories (such as massless
QCD)\ are less divergent than massive free-field theories!

In conclusion, the induced action for the conformal factor is convergent in
classically-conformal field theories, but not in the theories violating
conformality at the classical level.

\medskip

The reason why the convergence of the induced action is crucial for the
arguments of ref. \cite{athm} can be summarized as follows. In complete
generality, the bosonic terms of the classical action of a quantum field
theory should be positive-definite in the Euclidean framework for the
functional integral to make sense. The fermionic terms have a universal
form, in unitary theories. On physical grounds, we expect that, in a
physically acceptable theory, the bosonic part of the generating functional $%
\Gamma $ of the one-particle irreducible diagrams, which we call the quantum
action, be bounded from below. Shifting $\Gamma $ of a constant amount, we
can conveniently say that the quantum action is positive definite. Since,
however, the coupling constants run, it is more precise to say that the
quantum action $\Gamma $ is positive-definite throughout the RG flow if and
only if $\Gamma $ is positive-definite at a given energy.

Analysing a few simple examples, it is easy to get convinced that this
positivity property, which is spoiled in general by the regularization, is
recovered thanks to the renormalization procedure, and in particular the
running of coupling constants. Actually, renormalization can be viewed as
the unique algorithm which restores the mentioned positivity property by
means of local couterterms.

These considerations apply to the dependence of $\Gamma $ on
the dynamical fields. In the presence
of external fields, the positivity property is in general violated,
unless new dynamical parameters are introduced, associated with the
couplings to the external fields. The only possibility for the positivity
property to hold, in the presence of external sources, without adding new
parameters to the theory, is that the induced action be by itself convergent
in the presence of those sources. This happens if the sources are
coupled to evanescent operators.

The conformal factor $\phi$ is the external field coupled to the trace $%
\Theta$. $\Gamma[\phi]$ is assured to be convergent in classically-conformal
quantum field theories, by the very evanescence of $\Theta$. Instead, $%
\Gamma[\phi]$ is not convergent in the presence of masses or
super-rinormalizable interactions. For this reason, the positivity property
stated above applies only to the classically-conformal quantum field
theories.

The last step of the argument of \cite{athm}, which I do not repeat here,
was to prove that the mentioned positivity property implies the formula 
\begin{equation}
\Delta a={\frac{15}{2}}\pi ^{2}\int {\rm d}^{4}x\,|x|^{4}\,\langle \Theta
(x)\,\Theta (0)\rangle  \label{rog}
\end{equation}
in four dimensions. For the generalization of the argument to arbitrary even
dimensions the reader should refer to ref. \cite{at6d}.

I claim that the argument of \cite{athm} is a physical proof of the
irreversibility of the marginally relevant flows and of formula (\ref{rog}).

A non-trivial by-product of the analysis is that the marginally relevant
and the strictly relevant flows have an intrinsically different nature. The
investigation of this paper is useful to
understand the nature of this difference better.

\subsection{Open questions}

It is still not known how to generalize the theoretical argument just
recalled when classical conformality is violated. The knowledge we have at
present in this domain is only empirical \cite{cea}. We know that the flow
invariant of \cite{athm} is proportional to $\Delta c=c_{{\rm UV}}-c_{{\rm IR%
}}$ and not $\Delta a=a_{{\rm UV}}-a_{{\rm IR}}$, in massive free-field
theories. A natural implication of this fact is the definition of ``$c=a$''
theories \cite{cea}, which have interesting properties in this context (see
below). Yet, several issues remain open. Some of the most important
questions are:

{\it i)} What is the universal expression of the flow invariant?

{\it ii)} Why is the flow invariant, which is equal to $\Delta a$ in
marginally relevant flows, equal to $\Delta c$ in massive free-field
theories? What is it in general?

{\it iii)} What happens when the theory contains several 
scales ($\mu$, masses,
super-rinormalizable parameters, etc.)? 
Should the various scales be
related to one another in a special way, defining ``the'' flow, or should
they remain arbitrary, the result not depending on their relative values?

I call ``parameter'' the coefficient $\lambda_a$ of the
deformation ${\cal L}\rightarrow{\cal L}+\lambda_a {\cal O}_a$ of the
lagrangian ${\cal L}$. A dimensionful parameter 
is a parameter with non-vanishing
classical dimension in units of mass. In this definition, a theory can have
many scales, one for every dimensioned parameter $\lambda_a$, plus
the dynamical scale $\mu$. For the purposes of this paper, it is
necessary to keep these scales distinguished.
I do not write, for example, $\lambda_a=\Lambda ^{d_a}g_a$, $d_a$ being
the classical dimension of $\lambda_a$, to define a unique scale $\Lambda$
and several dimensionless parameters $g_a$, unless a special mechanism, such
as the spontaneous symmetry breaking, provides such relationships.

The investigations of this paper answer some open questions and address the
other ones. While several properties are understood better, other new
features emerge. A universal expression of the flow invariant is 
suggested by the results. This answers question $i$ and corrects a
previously proposed expression (see \cite{athm}, section 2.3), in the
presence of improvement terms for the stress tensor. The final formula is (%
\ref{final}).

The issue of universality of this flow invariant stands as follows. There is
evidence to claim universality in the following two subclasses of quantum
field theories:

a) marginally relevant flows;

b) relevant flows with $\Delta a=\Delta c$, in particular flows connecting
conformal fixed points with $c=a$.

In these cases, problem $ii$) does not show up. But the flow invariant is
expected to be useful also when $\Delta a\neq \Delta c$. I explain how,
after describing the flow invariant more precisely.

The formula of the flow invariant is made of three ingredients. First, a
flow integral of the type $\int {\rm d}^nx\, |x|^n\, \langle \Theta
(x)\,\Theta(0)\rangle$. As a second step, this integral is minimized over
the space of improved stress tensors. The result of this minimization will
still be called ``flow integral'', with an abuse of language, although in
reality it is a combination of flow integrals (see (\ref{sigma2})). Finally,
the formula of the flow invariant is made of a minimization in the space of
trajectories relating the dimensioned parameters of the theory.

\begin{figure}[tbp]
\begin{center}
\let\picnaturalsize=N 
\ifx\nopictures Y\else{\ifx\epsfloaded Y\else\fi
\global\let\epsfloaded=Y \centerline{\ifx\picnaturalsize N\epsfxsize
2.0in\fi \epsfbox{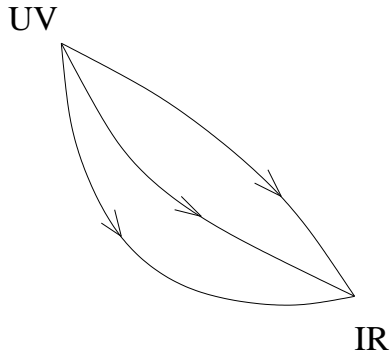}}}\fi
\par
\end{center}
\caption{Different flows connecting the same fixed points.}
\end{figure}

The analysis of the results suggests a non-trivial answer to question $iii$.
In the presence of several scales the value of the flow integral does depend
on the relations among them. The predicted value of the
invariant (which is $\Delta c$, in our cases) is the minimum of the flow
integral over all trajectories in the space of the dimensioned parameters of
the theory. Assuming that the dimensioned parameters of the theory are the
dynamical scale $\mu $ and some masses, or super-rinormalizable parameters, 
$m_{1},\ldots ,m_{k}$, a trajectory is, for example, a
set of functions $m_{1}(\mu ),\ldots ,m_{k}(\mu )$. Let $\bar{m}_{1}(\mu
),\ldots ,\bar{m}_{k}(\mu )$ denote the trajectory which minimises the flow
integral. Along $\bar{m}_{1}(\mu ),\ldots ,\bar{m}_{k}(\mu )$, the value of
the flow integral equals $\Delta c$. Along all other trajectories, it is
bigger than $\Delta c$.

Observe that the functions $m(\mu )$, or $\bar{m}(\mu )$ do not refer to the
running of the masses, but fix the reference values of the masses, which are
no longer
arbitrary, but depend themselves 
on the dynamical scale $\mu $. The complete $\mu $
dependence of the masses will be the result of the combination of the usual
running plus this relation. For example, the special trajectory $\bar{m}(\mu
)$ might be used to relate the value of the Higgs vev in the Standard Model
to the dynamical scale $\mu $ at a conventional reference energy. In the
special trajectory $\bar{m}(\mu )$, the Higgs mass is unambiguously fixed in
terms of $\Lambda _{{\rm QCD}}$. In these considerations, I am assuming 
that the Standard Model interpolates between well-defined UV and IR fixed 
points.

In conclusion, 
the use of the flow invariant in quantum field theories with  $\Delta
a\neq \Delta c$ is that it selects, by means of a minimum principle, 
priviledged flows among the set of flows connecting the same fixed points. 

We will find no clue to solve mystery $ii$. The investigation of this
challenging problem is left for the future. It might be useful, for this
purpose, to reconsider the issues studied here from the point of view of the
exact renormalization-group flow approach \'{a} la Wilson. 
I expect that, in the wilsonian
framework, the difference between marginally relevant and
truly relevant flows can be re-interpreted in the context of a
deeper and more complete understanding of the problem. In this paper, I use
only techniques of (resummed) perturbation theory.

Let us mention once again that in two dimensions the situation is much
simpler, because the flow invariant (\ref{fl2d}) does not depend of the
dimensioned parameters of the theory and there is a unique central charge.
The subclass of four-dimensional theories which best share the
properties of two-dimensional theories is the subclass 
with $\Delta c=\Delta a$ \cite{cea}.

\subsection{Applications and physical predictions}

The mentioned minimizations, and other properties studied in the paper,
suggest some new minimum principles, which might be important properties of
quantum field theory. For example, they might help us reformulating quantum
field theory in a framework where the renormalization-group trajectory
connecting two fixed points is obtained by minimising some action (see the
introduction of ref. \cite{high}). The idea is that a more sophisticated
formulation could better help us solve the open issues of quantum field
theory.

The investigation of flow invariants and quantum irreversibility, which is
a pretty theoretical research domain of quantum field theory,
allows us to make various types of quantitative predictions. For example, in
massless QCD, which is classically conformally invariant, $\Delta a$ should
be equal to 
\[
\Delta a={\frac{15}{2}}\pi ^{2}\int {\rm d}^{4}x\,|x|^{4}\,\langle \Theta
(x)\,\Theta (0)\rangle =62(N_{c}^{2}-1)+11N_{c}N_{f}-N_{f}^{2}+1, 
\]
where $N_{c}$ is the number of colours and $N_{f}$ is the number of quarks
in the fundamental representation. Lattice simulations in massless QCD with
fermions are difficult to perform, but not impossible. It would be extremely
interesting to test the prediction written above, because it crucially
depends of the field content of the IR limit of the theory. This check would
give an indirect evidence that the low-energy limit of the theory contains
just $N_{f}^{2}-1$ massless pions, as we expect. Other non-perturbative
checks of the formula for $\Delta a$ are possible in the context of the
AdS/CFT correspondence \cite{noi4}, but the computation is involved. The
simple flows of ref. \cite{loro4} are a good laboratory for this test.

Other types of physical predictions are available. The minimum principle
found here, which determines the priviledged trajectory $\bar{m}(\mu )$,
might have an important physical meaning, besides the mathematical one.
Suppose that the theories of nature are so constrained that they all lie on
the special trajectories $\bar{m}(\mu )$. Then, all the dimensioned physical
parameters are uniquely fixed in terms of the dynamical scale $\mu $, in all
the theories of nature, and we can relate the mass of the Higgs particle to $%
\mu $ and therefore predict the Higgs mass. This calculation is complicated
by the fact that it involves an integral throughout the flow. Because of its
intrinsically non-perturbative nature, this relation might generate very
large or very small numbers, possibly relevant for the hierarchy problem.
Work is in progress in this direction.

\subsection{Organization of the paper}

In section 2 I motivate the use of non-unitary theories for our
investigations. The precise formula for the flow invariant in the presence
of improvement terms for the stress-tensor will be constructed in two
successive steps. First, in section 3, I recall the candidate formula
proposed in sect. 2.3 of \cite{athm} and explain its properties. Then, in
sections 4, 5 and 6 I present checks in various higher-derivative models. In
section 4 I consider an example where improvement terms are irrelevant, but
the two fixed points are connected by a one-parameter family of
trajectories. The flow invariant does not depend on the trajectory and is
equal to $\Delta c$, as expected. In sections 5 and 6 I present checks where
the improvement terms are crucial. I show that the unimproved formula of the
flow invariant does not work, and that the candidate generalized formula
gives a result ``close'' to the prediction. The discrepancy is resolved in
section 7, where the correct formula of the flow invariant is found, and
associated with a minimum principle in the space of improvement terms. It is
then shown that another minimum principle (the one described 
above) defines a priviledged trajectory connecting the two conformal fixed
points. Along this priviledged trajectory the value of the flow integral is
equal to $\Delta c$, in agreement with the prediction, while along any other
trajectory it is greater than $\Delta c$. The second minimum principle
is then discussed in detail. Several calculations
are done numerically, but the final form of the flow invariant is simpler
than the first proposal and its value can be worked out exactly in many
cases.

\section{The use of non-unitary theories}

\setcounter{equation}{0}

As a laboratory, I study several higher-derivative theories. These
theories are not physical because not unitary. Nevertheless, they are useful
for the investigation of a variety of general properties of quantum field
theory. The idea is that the flow invariant is meaningful in all
mathematically well-defined theories, even if non-unitary, and that a lot
can be learned by enlarging the set of theories on which we can work. By
``mathematically well-defined theories'' I mean renormalizable theories such
that the bosonic contributions to the action are positive-definite in the
Euclidean framework. Conditions for the fermionic contributions are more
tricky and will be discussed later.

Mathematically well-defined theories are not necessarily unitary and even if
the bosonic part of the action is positive definite, they can violate
reflection positivity. It is easy to write negative-definite two-point
correlators and often, the central charges $a$ and $c$ are themselves
negative \cite{spin2}. Positive-definitness of the action assures, in
particular, that the propagators have no poles in the Euclidean framework,
so that Feynman diagrams are well-defined.

For the moment, I assume that the fermions, if present, have the Dirac
action $\bar\psi (D\!\!\!\!\slash+m(x))\psi$, where $m(x)$ is real and
eventually field dependent. It is not straightforward to set conditions for
acceptable non-unitary theories containing fermions. A specific model will
be analysed in the paper.

Reflection positivity is crucial in investigations about
the $c$-theorem \cite{zamolo}. On the other hand, reflection positivity is
less crucial in investigations about flow invariants.

Typically, in unitary theories a flow invariant is the integral of a
positive function along the flow. The positive integrand is a two-point
correlator. More generally, the flow invariant can be a positive combination
of integrals of this type. Consequently, the value of a flow invariant is
positive in unitary theories. By definition, its value does not depend on
the path connecting the two fixed points, but only on the fixed points
themselves. Moreover, it is additive. Concluding, the value of a flow
invariant has the form 
\[
\Delta_{{\rm UV}}-\Delta_{{\rm IR}}, 
\]
for some quantity $\Delta$, which we can call ``central charge'' of the
conformal fixed point. In various cases it can be identified with $a$ or $c$%
. Conformal field theories might have more than one central charge and there
might exist several flow invariants.

Under the conditions just stated, a unitary theory will satisfy the
inequality 
\begin{equation}
\Delta_{{\rm UV}}-\Delta_{{\rm IR}}\geq 0,  \label{ineq}
\end{equation}
or ``$c$-theorem''. A non-unitary theory is allowed to violate this
inequality. The integrand of the flow invariant, which is, as we have said,
a two-point correlator, can be negative.

Since, however, our interest is not to test an inequality such as (\ref{ineq}%
), but flow invariance, and to possibly classify all flow invariants, it
does not matter to us whether the inequality (\ref{ineq}) is satisfied or
not. The tests
presented here confirm that the flow invariant is generically well-defined
in the matematically well-defined theories, even if they are non-unitary,
modulo some conditions on the higher-derivative fermions.

Renormalizability puts formidable constraints on the set of allowed physical
theories, which are often not enough numerous, or not sufficiently simple,
to make ``quantum field theoretical experiments'' of the type that we are
going to present here. For example, improvement terms for the stress tensor
are important in theories containing scalar fields, such as the Standard
Model. However, the improvement terms affect the flow invariant above four
loops in the $\lambda\varphi^4$ theory \cite{athm} and it is often difficult
to perform calculations to high orders in perturbation theory. Extending the
set of theories to the mathematically, but not necessarily physically,
well-defined ones, in the sense specified above, we can enlarge the
laboratory of theories enormously, in arbitrary dimensions. This laboratory
includes also classically conformal higher-spin theories \cite{spin2}, which
sometimes have conformal windows and can be coupled to external gravity. A
variety of new checks and ``theoretical experiments'' are available and
calculations are easier.

Moreover, higher-derivative theories and other non-unitary theories, have a
number of non-trivial mathematical applications. For example, they can be
studied to investigate the properties of the ``pondered'' Euler density
constructed in \cite{at6d}, or the special invariant appearing in the trace
anomaly of the so-called ``$c=a$ theories'', constructed in \cite{cea}. 
In six dimensions this special invariant matches with a
particular combination pointed out by Bonora, Pasti and Bregola in ref. \cite
{bonora}, and in higher even dimensions agrees with the Henningson-Skenderis
construction \cite{HS}. Quantum field theory is the most powerful algorithm
to study these mathematical properties.

The results presented here and in \cite{athm,at6d,cea} suggest that quantum
field theory has a number of unforeseen, interesting properties,
which can be investigated exactly, without
restricting to particularly symmetric theories. This, 
in spite of the usual lore
that the open problems of quantum field theory are too difficult.

The claimed difficulty has often been advocated as a motivation to state
that exact properties of quantum field theory should be studied using more
powerful, ``non-perturbative'' methods, such as those suggested by string
theory. While it is certainly true that the string theoretical methods can
improve and extend our knowledge of quantum field theory, it is also
doubtless that their power is limited. Typically, these methods cover a very
special subclass of quantum field theories, having some peculiar symmetry
or related to one another by some
sort of duality. Not unfrequently, the physically interesting theories are
excluded. For example, the AdS/CFT correspondence is useful to study
conformal theories with $c=a$ \cite{HS} and flows interpolating between them 
\cite{gira}. The $c=a$ conformal field theories and flows are certainly
interesting \cite{cea}, but physically relevant theories (QCD, Standard
Model) do not belong to this class \cite{high}.

Instead, the spirit of the investigation presented here is that the
knowledge coming from the analysis of a variety of physically uninteresting
theories, then applies also to the physically interesting ones. For example,
we are going to find the precise, general formula of the flow invariant in
the presence of improvement terms for the stress tensor, which can be
applied to the Standard Model. We could not infer this formula directly from
the Standard Model.

In conclusion, the investigation of the mathematically well-defined quantum
field theories is well motivated.

\section{First step towards the universal flow invariant}

\setcounter{equation}{0}

The central charge $c_n$ in $n$ dimensions is normalized so that it equals
one for a real free scalar field. In even dimensions, we normalize the
central charge $a_n$ so that the $c=a$ theories defined in ref. \cite{cea}
are those satisfying the condition $c_n=a_n$. The two relevant terms of the
trace anomaly in external gravity read 
\[
\Theta={\frac{\left({\frac{n}{2}}\right)!}{(4\pi)^{n/2}\, (n+1)!}} \left[{%
\frac{a_n}{2^{n/2-1}\, n}}\,{\rm G}_n-c_n{\frac{n-2}{4(n-3)}}W\Box ^{n/2-2}W
+\cdots, \right] 
\]
where 
\[
{\rm G}_n={\rm G}_{n}=(-1)^{\frac{n}{2}} \varepsilon _{\mu _{1}\nu
_{1}\cdots \mu _{\frac{n}{2}}\nu _{\frac{n}{2}}}\varepsilon ^{\alpha
_{1}\beta _{1}\cdots \alpha _{\frac{n}{2}}\beta _{\frac{n}{2}}}\prod_{i=1}^{%
\frac{n}{2}}R_{\alpha _{i}\beta _{i}}^{\mu _{i}\nu _{i}} 
\]
is the Gauss-Bonnet integrand and $W$ is the Weyl tensor. Observe that the $%
a_n$-normalization chosen in \cite{at6d,cea} was such that $\Theta=a_n \, 
{\rm G_n}+\cdots$.

In both even and odd dimensions, the central charge $c$ can be defined by
the stress-tensor two-point function. Using the notation of ref. \cite{iera}%
, we have: 
\begin{equation}
\langle T_{\mu \nu }(x)\,T_{\rho \sigma }(0)\rangle = c_n{\frac{(n/2)!
(n/2-1)!}{(2\pi)^n (n+1)!}} {\prod }_{\mu \nu ,\rho \sigma
}^{(2)}\Box^{n/2-2} \left( {\frac{1}{|x|^{n}}}\right),  \label{twop}
\end{equation}
where the spin-2 projection operator ${\prod }^{(2)}$ is given by 
\[
{\prod }_{\mu \nu ,\rho \sigma }^{(2)}={\frac{1}{2}} (\pi_{\mu\rho}\pi_{\nu%
\sigma}+\pi_{\mu\sigma}\pi_{\nu\rho}) -{\frac{1}{n-1}}\pi_{\mu\nu}\pi_{\rho%
\sigma},~~~~~~~~~ \pi_{\mu\nu}=\partial_\mu\partial_\nu-\Box\delta_{\mu\nu}. 
\]
In odd dimensions, we understand that 
\[
\Box^{n/2-2}\left({\frac{1}{|x|^n}}\right)= {\frac{2^{n-3} (n-3)!}{%
(n-2)\,|x|^{2n-4}}}. 
\]
On the other hand, a consistent definition of the central charge $a_n$ in
odd dimensions is still missing.

The flow invariant proposed in sect. 2.3 of \cite{athm}, generalized to many
improvement terms, reads 
\begin{equation}
\Sigma_n=\int {\rm d}^{n}x \,|x|^n {\frac{ \det\left[\matrix{\langle
\Theta(x) \, \Theta(0)\rangle &\langle \Theta(x)\, {\cal O}_j(0)\rangle \cr
\langle {\cal O}_i(x)\, \Theta(0)\rangle & \langle {\cal O}_i(x)\, {\cal
O}_j(0)\rangle}\right] }{\det\left[\langle {\cal O}_i(x)\, {\cal O}%
_j(0)\rangle \right]}} \equiv\int {\rm d}^{n}x \,|x|^n\, \sigma(x).
\label{sigma}
\end{equation}
Here ${\cal O}_i$, $i=1,\ldots k$, are the traces of the improvement terms
for the stress tensor. This formula was suggested from considerations about
the scheme independence and flow invariance in the $\varphi^4$-theory in
four dimensions, which has a single improvement term ($k=1$).

The improvement terms ${\cal O}_i$ are classified as follows. We assume that
the stress tensor $T_{\mu\nu}$ is traceless at the critical points. We
consider all local, dimension $n$, symmetric, identically conserved
operators $\Delta T_{\mu_\nu}$, which vanish at the critical points.
Clearly, the ``improved'' operators of the form $T_{\mu\nu}+\Delta
T_{\mu_\nu}$ are equally acceptable stress tensors.

Translation invariance implies that $\partial_\mu T_{\mu\nu}$ is finite, but
not necessarily that $T_{\mu\nu}$ is finite \cite{itzykson}. It can be
inferred that $T_{\mu\nu}$ is finite only if there exists no identically
conserved local operator $\Delta T_{\mu\nu}$. In the $\lambda\varphi^4$%
-theory in four dimensions such an operator exists, and is equal to $%
h(\lambda)(\partial_\mu\partial_\nu-\delta_{\mu\nu}\Box)(\varphi^2)$, where $%
h(\lambda)$ is an arbitrary function of $\lambda$, vanishing at the fixed
points. The bare and renormalized stress tensors are related by 
\[
T_{\mu\nu}^{R}=T_{\mu\nu}^{B}+A(\lambda)
(\partial_\mu\partial_\nu-\delta_{\mu\nu}\Box)(\varphi^2)^{B} 
\]
where $A$ is possibly divergent. Similarly, we have, for the traces, $%
\Theta^{R}=\Theta^{B}-3A\Box(\varphi^2)^{B}$. We have therefore ${\cal O}%
=\Box(\varphi^2)$ and 
\begin{equation}
\sigma=\langle \Theta\,\,\Theta\rangle -{\frac{\left(\,\langle\Theta\,\,\Box%
\varphi^2\rangle\,\right)^2 }{\langle\Box\varphi^2\,\,\Box \varphi^2\rangle}}
\label{sigmascalar}
\end{equation}

Observe that, instead, the massive free scalar field admits no improvement
term. The operator $\Box(\varphi^2)$ cannot be multiplied by a dimensionless
function, vanishing at criticality, because there exists no dimensionless
parameter in the theory. If we make a redefinition of the form $%
T_{\mu\nu}\rightarrow T_{\mu\nu}-\alpha/3 \,
(\partial_\mu\partial_\nu-\delta_{\mu\nu}\Box)(\varphi^2)$ (and consequently 
$\Theta\rightarrow \Theta+\alpha \Box(\varphi^2)$), then $\alpha$ has to be
constant. That means, however, that the shift $\alpha \Box(\varphi^2)$
cannot vanish at the critical points, where, instead, the condition $%
\Theta=0 $ defines the trace (and so the stress tensor) uniquely. Therefore,
the stress tensor is unique also off-criticality for the massive free scalar
field.

As a third example, useful for the investigation of this paper, let us
consider a higher-derivative massive free scalar field $\varphi$ with
lagrangian $1/2\,[(\Box \varphi)^2+m^4\varphi^2]$. In this theory, $\varphi$
has dimension $(n-4)/2$. We can set $\Delta T_{\mu\nu}=m^2
(\partial_\mu\partial_\nu-\delta_{\mu\nu}\Box)(\varphi^2)$. The coefficient $%
m^2$ assures that the improvement term disappears at criticality. We have
therefore ${\cal O}=m^2\Box(\varphi^2)$. Nevertheless, we will see below
that the flow invariant does not depend on the coefficient of the
improvement operator, and so we can just take ${\cal O}=\Box(\varphi^2)$.

\medskip

The integrand $\sigma(x)$, and so $\Sigma_n$, are independent of the choice
of $T_{\mu\nu}$ among the set of improved stress tensors $T_{\mu\nu}+\Delta
T_{\mu\nu}$. On $\Theta$, this means invariance under the transformation 
\begin{equation}
\Theta\rightarrow\Theta+a_i{\cal O}_i, \qquad {\cal O}_i\rightarrow B_{ij}%
{\cal O}_j,  \label{inva}
\end{equation}
$a_j$ and $B_{ij}$ being arbitrary constants.

Equivalently, the meaning of the invariance (\ref{inva}) is that $\Sigma_n$
and $\sigma(x)$ are independent of the coupling of the theory to the
gravitational background. At the critical points the non-minimal couplings
are fixed uniquely by conformal invariance, but at intermediate energies
they affect the stress tensor and $\Theta$. The
non-minimal couplings are in one-to-one
correspondence with the improvement terms of the stress tensor. For example,
in the $\lambda\varphi^4$-theory embedded in external gravity, the
non-minimal coupling is $R\varphi^2$. By definition, the non-minimal
couplings disappear in the flat limit. Therefore, the flatspace
 theory should be
insensitive to them.

The integrand $\sigma(x)$ has other relevant invariance properties. In
particular, it was shown in sect. 2.3 of \cite{athm} that (\ref{sigmascalar}%
) is scheme independent and that it dependes on the running coupling $%
\lambda(1/|x|)$, but not on the reference value $\lambda(\mu)$ of the
coupling. The origin of the possible scheme dependende is the
renormalization mixing between the stress tensor and its improvement term.
The scheme dependence cancels out in the combination $\sigma(x)$.

The matrix $B$ in (\ref{inva}) is the most general real matrix, because
there is no canonical way of normalizing the improvement operators. For
example, (\ref{sigmascalar}) is independent on the normalization of the
operator $\Box\varphi^2$.

The invariance (\ref{inva}) can be proved as follows. In the first case, let 
$a_i=0$. It is easy to see that the redefinitions ${\cal O}_i^{\prime}=B_{ij}%
{\cal O}_j$ produce 
\[
\sigma^{\prime}={\frac{\det\left[\left(\matrix{1 & 0 \cr 0 & B}\right) \left(%
\matrix{\langle \Theta \,\, \Theta\rangle &\langle \Theta\, {\cal O}_j\rangle \cr   
\langle {\cal O}_i\, \Theta\rangle & \langle {\cal O}_i\, 
{\cal O}_j\rangle}\right) \left(\matrix{1 & 0 \cr 0 & B^t}\right)\right] }{%
(\det B)^2 \, \det\left[\langle {\cal O}_i\, {\cal O}_j\rangle \right]}}
=\sigma, 
\]
since the determinants of the matrix $B$ simplify. Here $B^t$ denotes the
transpose of $B$.

Now, let us take $B=0$ and $a_i\neq 0$. To prove invariance, we rewrite $%
\sigma$ by expanding the numerator determinant along the first row and the
first column: 
\[
\sigma=\langle \Theta \,\, \Theta\rangle+ \sum_{i,j=1}^k(-1)^{i+j+1}\langle
\Theta\, {\cal O}_j\rangle \langle {\cal O}_i\, \Theta\rangle \,{\frac{\det
N_{ij}}{\det N}}, 
\]
where $N$ denotes the matrix with entries $\langle {\cal O}_i\, {\cal O}%
_j\rangle$ and $N_{ij}$ is the minor obtained by suppressing the $i^{th}$
row and the $j^{th}$ column. Defining $\Theta^{\prime}=\Theta+a_i{\cal O}_i$
we can readily show $\sigma^{\prime}=\sigma$ by means of the well known
identity 
\[
\sum_{i=1}^k \langle {\cal O}_k\, {\cal O}_i\rangle \, \det N_{ij}\,
(-1)^{i+j+1}=-\delta_{kj}\, \det N. 
\]

\medskip

We have recalled in the introduction that the flow invariant is proportional
to $\Delta a$ in marginally relevant flows. The proportionality coefficient
can be read from \cite{at6d}. In the normalization used here, we have 
\[
\Sigma_n=\Delta a_n\, {\frac{\Gamma(n/2+1)}{\pi^{n/2}\, (n+1)}}. 
\]
Instead, we know that the flow invariant is proportional to $\Delta
c_n=c_{n\, {\rm UV}}-c_{n\, {\rm IR}}$ is unitary, massive free-field
theories. The proportionality coefficient can be read from \cite{cea}: 
\begin{equation}
\Sigma_n=\Delta c_n\, {\frac{\Gamma(n/2+1)}{\pi^{n/2}\, (n+1)}}.
\label{predic1}
\end{equation}
The theories studied in this paper are free higher-derivative massive
bosonic and fermionic theories. We therefore expect that the prediction (\ref
{predic1}) applies to our case.

However, the results do not confirm this prediction. Later we will see that
this failure has two reasons.

First, the proposed formula for the flow invariant is not correct in the
presence of improvement terms for the stress tensor. The point is that the
flow invariant is not uniquely fixed by the symmetry (\ref{inva}). The
integral $\Sigma_n$ is invariant under an enlarged symmetry, of the same
form as (\ref{inva}), but with point-dependent $a_i$ and $B_{ij}$. This is
an unnecessary requirement, since it is associated with non-local
non-minimal coulings to external gravity. The flow invariant has certainly
to be invariant under (\ref{inva}), with constant $a_i$ and $B_{ij}$, but it
need not be the integral of an invariant integrand $\sigma(x)$.

Second, the correct flow integral is not always equal to the right-hand side
of (\ref{predic1}), but this happens in a special trajectory connecting the
UV and IR fixed points.

For pedagogical reasons, and taking into account of the empirical nature of
the investigations of this paper, I present the results as I have found
them, starting from the more intuitive, although incorrect, formulations of
the prediction and the flow invariant, and letting the correct formulas
emerge from the analysis of the results.

\section{First test of the prediction}

\setcounter{equation}{0}

As a first test, we consider a higher-derivative theory of fermions with
lagrangian 
\[
{\cal L}=\bar\psi(\partial\!\!\!\slash-m_1)(\partial\!\!\!\slash+m_2)\psi. 
\]
This theory has no improvement term. We have a one-parameter family of
relevant flows connecting the same pair of conformal fixed points, the
parameter being $m_1/m_2$.

The stress tensor reads 
\begin{eqnarray*}
T_{\mu\nu}&=&{\frac{1}{4}}\left[\bar\psi\gamma_\mu \overleftrightarrow{%
\partial_\nu}(\partial\!\!\!\slash+m_2)\psi+ \bar\psi\gamma_\nu 
\overleftrightarrow{\partial_\mu}(\partial\!\!\!\slash+m_2)\psi- \bar\psi(%
\overleftarrow{\partial\!\!\!\slash}+m_1)\gamma_\mu \overleftrightarrow{%
\partial_\nu}\psi- \bar\psi(\overleftarrow{\partial\!\!\!\slash}%
+m_1)\gamma_\nu \overleftrightarrow{\partial_\mu}\psi \right]  \nonumber \\
&-&{\frac{1}{2}}\delta_{\mu\nu}\left(\bar\psi(\partial\!\!\!\slash-m_1)
(\partial\!\!\!\slash+m_2)\psi+ \bar\psi(\overleftarrow{\partial\!\!\!\slash}
-m_1)(\overleftarrow{\partial\!\!\!\slash}+m_2)\psi\right).
\end{eqnarray*}
where we have kept the terms proportional to the field equations.

The trace is 
\[
\Theta=-\bar\psi\overleftarrow{\partial\!\!\!\slash}\partial\!\!\!\slash%
\psi+ m_1m_2\bar\psi\psi. 
\]
Observe that the trace is non-vanishing in the massless limit. Nevertheless,
the theory is conformal at $m_1=m_2=0$, since the operator $\bar\psi%
\overleftarrow{\partial\!\!\!\slash}\partial\!\!\!\slash\psi$ has a
vanishing two-point function: $\langle\Theta(x)\, \Theta(0)\rangle=0$ at $%
x\neq 0$ and $m_1=m_2=0$. This can be shown immediately using the massless
field equations. Moreover, the correlator $\langle T_{\mu\nu}(x)\,
\Theta(0)\rangle$ is also vanishing at $x\neq 0$, since each term in the
stress tensor contains either $\partial \!\!\!\slash\psi$ or $\bar\psi%
\overleftarrow{\partial\!\!\!\slash}$. This is a peculiarity of
higher-derivative theories, which violate reflection positivity. A
nonvanishing operator can have a vanishing two-point function. In a physical
reduction of the theory, admitting that it exists, such operators should be
projected away, presumably in a cohomological sense.

At $m_1=m_2=0$ the central charge $c_n$ is positive and twice the one of
ordinary fermions: 
\[
c_n=2^{[n/2]}(n-1), 
\]
where $[n/2]$ denotes the integral part of $n/2$. Our prediction reads
therefore 
\begin{equation}
\Sigma_n=\int {\rm d}^{n}x \,|x|^n \langle \Theta(x) \, \Theta(0)\rangle= {%
\frac{2^{[n/2]}(n-1) \Gamma(n/2+1)}{\pi^{n/2} (n+1)}}.  \label{preda}
\end{equation}

I have checked this prediction in six, eight and ten dimensions, for various
values of $m_1/m_2$, performing the integral $\Sigma_n$ has numericaly in
the momentum space. When $m_1=m_2=m$ it is easy to compute the flow integral
exactly. We have 
\[
\langle \Theta(x) \, \Theta(0)\rangle= 2^{[n/2]+1}m^2(G_n^{\prime\,
2}(x)-m^2 G_n^2(x)), ~~~~~~~G_n(x)=-{\frac{1}{2\pi}}\left({\frac{m}{2\pi|x|}}%
\right)^{n/2-1}K_{n/2-1}(m|x|), 
\]
$K$ denoting the modified Bessel function, and 
\[
\int {\rm d}^{n}x \,|x|^n \, m^2 (G_n^{\prime \, 2}-m^2 G_n^2)= {\frac{%
(n-1)\Gamma(n/2+1)}{2 \,\pi^{n/2} (n+1)}}, 
\]
which, inserted into the left-hand side of (\ref{preda}), produces the
correct result. The numerical results confirm the prediction for arbitrary
values of $m_1/m_2$.

\section{Second test of the prediction}

\setcounter{equation}{0}

We now consider the scalar theory defined by the lagrangian 
\[
{\cal L}={\frac{1}{2}}\left[(\Box\varphi)^2+\beta
m^2(\partial_\alpha\varphi)^2 +m^4\varphi^2\right]. 
\]
The stress tensor can be fixed in the following way. At $m=0$ $T_{\mu\nu}$
is conserved and traceless. There is a unique expression, up to an overall
constant, satisfying this condition. In six dimensions this expression was
found in ref. \cite{spin2}: 
\begin{eqnarray*}
T_{\mu \nu } &=&h\left\{ \frac{3}{4}\partial _{\mu }\partial _{\alpha
}\varphi\, \partial _{\nu }\partial _{\alpha }\varphi -\frac{3}{2}\Box
\varphi\, \partial _{\mu }\partial _{\nu }\varphi +\partial _{\nu }\Box
\varphi\, \partial _{\mu }\varphi +\partial _{\mu }\Box \varphi\, \partial
_{\nu }\varphi -\frac{1}{2}\partial _{\mu }\partial _{\nu }\partial _{\alpha
}\varphi\, \partial _{\alpha }\varphi \right. \\
&&-\frac{1}{4}\varphi\, \Box \partial _{\mu }\partial _{\nu }\varphi +\left.
\delta _{\mu \nu }\left[ -\frac{1}{4}\partial _{\alpha }\Box\, \varphi
\partial _{\alpha }\varphi -\frac{1}{8}\left( \partial _{\alpha }\partial
_{\beta }\varphi \right) ^{2}+\frac{1}{4}\left( \Box \varphi \right)
^{2}\right] \right\}
\end{eqnarray*}
To fix the overall constant, it is sufficient to inspect the stress tensor
up to total derivatives: 
\[
T_{\mu\nu}=-{\frac{5}{2}}h\varphi\, \partial_\mu\partial_\nu\Box\varphi +%
{\rm tot.}~{\rm ders.} ={\frac{2}{\sqrt{g}}}\left. {\frac{\delta}{\delta
g^{\mu\nu}}}\int{\frac{1}{2}}\sqrt{g} (\partial_\rho\partial_\sigma\varphi)(%
\partial_\alpha\partial_\beta\varphi)
g^{\rho\sigma}g^{\alpha\beta}\right|_{g_{\mu\nu}=\delta_{\mu\nu}}+ {\rm tot.}%
~{\rm ders.} 
\]
This gives $h=-4/5$. Finally, the contributions of the mass operators can be
straightforwardly calculated from the embedding in external gravitaty.

The stress-tensor two-point function, at $m=0$ in $n=6$, can be read from
ref. \cite{spin2}: 
\[
\langle T_{\mu \nu }(x)\,T_{\rho \sigma }(0)\rangle =-{\frac{1}{ 5376\pi ^{6}%
}}{\prod }_{\mu \nu ,\rho \sigma }^{(2)}\Box \left( {\frac{1}{|x|^{6}}}%
\right) =c_6{\frac{2 \cdot 3!}{2^6\cdot 7!\cdot\pi^6 }}{\prod }_{\mu \nu
,\rho \sigma }^{(2)}\Box \left( {\frac{1}{|x|^{6}}}\right). 
\]
This gives the central charge 
\[
c_6=-{5}. 
\]
The negative sign of the central charge signals the presence of ghosts in
the theory.

The prediction (\ref{predic1}) finally reads 
\[
\Sigma_6=-{\frac{30}{7\pi^3}}. 
\]

We can generalize the calculations and the prediction to arbitrary
dimensions. The stress tensor is easily found to be 
\begin{eqnarray*}
T_{\mu \nu } &=&- \frac{n+2}{2(n-1)}(\partial _{\nu }\Box \varphi \,
\partial_{\mu }\varphi +\partial _{\mu }\Box \varphi\, \partial _{\nu
}\varphi) + \frac{n(n+2)}{2(n-1)(n-2)}\Box \varphi \, \partial
_{\mu}\partial _{\nu }\varphi +\frac{2}{n-1}\partial _{\mu }\partial _{\nu
}\partial _{\alpha }\varphi \,\partial _{\alpha }\varphi \\
&&-{\frac{2n}{(n-1)(n-2)}} \partial _{\mu }\partial _{\alpha }\varphi\,
\partial _{\nu }\partial _{\alpha }\varphi +\frac{n-4}{2(n-1)}\varphi\, \Box
\partial _{\mu }\partial _{\nu }\varphi +\delta_{\mu \nu }\left[ \frac{1}{n-1%
}\partial _{\alpha }\Box \varphi\, \partial _{\alpha}\varphi \right. \\
&&\left. +\frac{2}{(n-1)(n-2)}\left( \partial _{\alpha }\partial _{\beta }
\varphi \right)^{2} -\frac{n+2}{2(n-1)(n-2)}\left( \Box \varphi \right) ^{2}
-{\frac{n-4}{2(n-1)}}\varphi\, \Box^2\varphi-{\frac{m^4}{2}}\varphi^2\right]
\\
&& +\beta m^2\left(\partial_\mu\varphi\,\partial_\nu\varphi- {\frac{%
\delta_{\mu\nu}}{2}}(\partial_\alpha\varphi)^2 \right)-\alpha
m^2(\partial_\mu\partial_\nu-\Box\delta_{\mu\nu}) (\varphi^2).
\end{eqnarray*}
The result can be derived also from the complete coupling to gravity, which
is known in this case. We have 
\begin{equation}
{\cal L}={\frac{1}{2}}\sqrt{g}\left(\varphi\Delta_4\varphi+ \beta
m^2(\partial_\mu\varphi)(\partial_\nu\varphi)g^{\mu\nu} +\alpha R m^2
\varphi^2+m^4 \varphi^2\right),  \label{uno}
\end{equation}
where (see for example \cite{johanna}) 
\begin{eqnarray}
\Delta_4&=&\nabla^2\nabla^2+\nabla_\mu\left[ {\frac{4}{n-2}}R^{\mu\nu}-{%
\frac{n^2-4n+8}{2(n-1)(n-2)}}g^{\mu\nu}R\right] \partial_\nu-{\frac{n-4}{%
4(n-1)}}\nabla^2R  \nonumber \\
&& -{\frac{n-4}{(n-2)^2}}R_{\mu\nu}R^{\mu\nu}+{\frac{(n-4)(n^3-4n^2+16n-16) 
}{16(n-1)^2(n-2)^2}}R^2.  \label{due}
\end{eqnarray}
The stress tensor can be obtained by direct differentiation of the
lagrangian embedded in the external gravitational field. The result agrees
with the previous calculation and ref. \cite{spin2}.

A lengthy calculation gives 
\[
\langle T_{\mu \nu }(x)\,T_{\rho \sigma }(0)\rangle =-{\frac{(n-4)(n^2-16)
\left[\Gamma(n/2-2)\right]^2}{2^{n+4}\pi^n(n^2-1)\Gamma(n-2) }}\,{\prod }%
_{\mu \nu ,\rho \sigma }^{(2)}\Box^{n/2-2} \left( {\frac{1}{|x|^{n}}}\right) 
\]
and therefore the central charge 
\[
c_n=-{\frac{2(n+4)}{n-2}}, 
\]
which agrees with the known results for $n=6$ and $n=4$. The value for $n=4$
can be found in \cite{spin2}, formula (2.4), where an additional factor
1/120 is included, in the more conventional four-dimensional normalization.
Here, instead, we have normalized $c$ to be always 1 for a free real scalar
field.

The prediction (\ref{predic1}) reads 
\begin{equation}
\Sigma_n=-{\frac{2(n+4)\Gamma(n/2+1)}{(n+1)(n-2)\,\pi^{n/2}}}.  \label{pred}
\end{equation}
We now proceed to check this prediction.

\medskip

The propagator is a convolution of two Bessel functions. Precisely, 
\begin{eqnarray*}
G_{n,r}(x)&\equiv& \int{\frac{{\rm d}^n p}{(2\pi)^n}}{\frac{{\rm e}^{ipx}}{%
(p^2+m^2\gamma_+^2) (p^2+m^2\gamma_-^2)}}={\frac{1}{(2\pi)^{n/2}\,x^{n-4}}}%
\int_0^\infty {\frac{t^{n/2}\,J_{n/2-1}(t)\, {\rm d}t}{(t^2+rm^2x^2)%
\left(t^2+{\frac{1}{r}}m^2x^2\right)}},
\end{eqnarray*}
where $\gamma_-=1/\gamma_+$ and $\beta=\gamma_+^2+\gamma_-^2$ and $%
\gamma_+^2=r$. The propagator can be written more explicitly in two classes
of cases: in odd dimensions, for any value of $r$, and in even dimensions,
for $r=1$. For example in three, five, seven and nine dimensions we have 
\begin{eqnarray*}
G_{3,r}(x)&=&{\frac{{\rm e}^{-m|x|\gamma_-}-{\rm e}^{-m|x|\gamma_+} }{4 \pi
m^2 |x|(\gamma_+^2-\gamma_-^2)}}~~~~~~~~~~~~~~~~~~~~~~~~~~~~
~~~~~~~~~~~~~\rightarrow {\frac{{\rm e}^{-m|x|}}{8\pi m}}, \\
G_{5,r}(x)&=&{\frac{(1+m|x|\gamma_-)\,{\rm e}^{-m|x|\gamma_-}
-(1+m|x|\gamma_+)\, {\rm e}^{-m|x|\gamma_+} }{8 \pi^2 m^2 |x|^3
(\gamma_+^2-\gamma_-^2)}}~~~~~~\rightarrow {\frac{{\rm e}^{-m|x|}}{16\pi^2
|x|}}, \\
G_{7,r}(x)&=&{\frac{(3+3m|x|\gamma_-+m^2|x|^2\gamma_-^2)\, {\rm e}%
^{-m|x|\gamma_-} -(3+3m|x|\gamma_++m^2|x|^2\gamma_+^2)\, {\rm e}%
^{-m|x|\gamma_+} }{16 \pi^3 m^2 |x|^5 (\gamma_+^2-\gamma_-^2)}} \\
&& ~~~~~~~~~~~~~~~~~~~~~~~~~~~~~~~~~~~~~~~~~~~~~~~~~~~~~~
~~~~~~~~~~~~~\rightarrow {\frac{(1+m|x|)\,{\rm e}^{-m|x|}}{32\pi^3 |x|^3}},
\\
G_{9,r}(x)&=&{\frac{(15 + 15m |x|\gamma_- + 6m^2 |x|^2 \gamma_-^2 + m^3
|x|^3\gamma_-^3)\, {\rm e}^{-m|x|\gamma_-}}{32\pi^4 m^2|x|^7
(\gamma_+^2-\gamma_-^2)}} \\
&-& {\frac{(15 + 15m |x|\gamma_+ + 6m^2|x|^2 \gamma_+^2 + m^3|x|^3
\gamma_+^3) \, {\rm e}^{-m|x|\gamma_+} }{32\pi^4 m^2|x|^7
(\gamma_+^2-\gamma_-^2)}} ~~\rightarrow {\frac{(3 + 3 m |x| + m^2 x^2)\, 
{\rm e}^{-m|x|}}{64 \pi^4 |x|^5}}.
\end{eqnarray*}
On the right hand sides of the arrows, the expressions of the propagators
for $r=\gamma_+=\gamma_-=1$ are reported.

For $r=1$ the function $G_{n,1}(x)$ can be written, in arbitrary dimensions,
by means of a modified Bessel function: 
\begin{equation}
G_{n,1}(x)={\frac{1}{2(2\pi)^{n/2}}}\left({\frac{m}{|x|}}\right)
^{n/2-2}K_{n/2-2}(m|x|).  \label{gn1}
\end{equation}

The trace of the stress tensor reads 
\[
\Theta=-2m^4\varphi^2-{\frac{n-2}{2}}\beta m^2(\partial_\alpha\varphi)^2 -{%
\frac{n-4}{2}}\beta m^2 \varphi\,\Box\varphi+\alpha(n-1)m^2\Box(\varphi^2), 
\]
using the field equations $\Box^2\varphi-\beta m^2\Box\varphi+m^4\varphi=0$.
Recalling that $\Sigma_n$ is invariant under the redefinition $%
\Theta\rightarrow \Theta+\gamma\Box(\varphi^2)$, with $\gamma$ arbitrary, we
can use the following simplified expression for $\Theta$: 
\[
\Theta^\prime=-2m^4\varphi^2+\beta m^2\varphi\,\Box\varphi. 
\]
Observe that this expression does not contain an explicit dependence on the
dimension.

We find 
\begin{eqnarray*}
\sigma(x)&=&\langle \Theta^{\prime}(x)\,\Theta^{\prime}(0)\rangle -{\frac{%
\left(\,\langle\Theta^{\prime}(x)\,\Box\varphi^2(0)\rangle\,\right)^2 }{%
\langle\Box\varphi^2(x)\,\Box \varphi^2(0)\rangle}} \\
&=& 8m^8G^2+\beta^2m^4\left(G\Box^2 G+(\Box G)^2\right) -8\beta m^6G\Box G
-2m^4{\frac{\left[\Box\left(2m^2 G^2-\beta G\Box G\right) \right]^2 }{\Box^2
G^2 }}.
\end{eqnarray*}

The numerical results are summarized in tables 1 and 2, where the ratio $R$
between the calculated value of the flow integral and the predicted value (%
\ref{pred}) is reported.

\begin{table}[tbp]
\begin{center}
\begin{tabular}{|cc|cc|cc|}
\hline
$n$ & $R$ & $n$ & $R$ & $n$ & $R$ \\ \hline
~4 & sing. ~ & ~ 10 & 1.14836~ & ~ 70 & 1.00304~ \\ 
~5 & sing. ~ & ~ 12 & 1.09797~ & ~ 100 & 1.00152~ \\ 
~6 & 2.12370 ~ & ~ 15 & 1.06120~ & ~ 150 & 1.00069~ \\ 
~7 & 1.41636 ~ & ~ 20 & 1.03440~ & ~ 200 & 1.00039~ \\ 
~8 & 1.26673 ~ & ~ 30 & 1.01564~ & ~ 300 & 1.00017~ \\ 
~9 & 1.19275 ~ & ~ 45 & 1.00716~ & ~ 500 & 1.00006~ \\ \hline
\end{tabular}
\end{center}
\caption{Ratio between calculated and predicted values of $\Sigma_n$ for $%
r=1 $ in various dimensions $n$.}
\end{table}

In table 1 we show the results for $r=1$. In 4 and 5 dimensions, the
interand $\sigma(x)$ is singular, and $\Sigma$ is ill-defined. This might
either mean that the expression of the flow invariant is not correct, or
that it does not make sense to apply the theory of flow invariants to
non-unitary theories. The integral $\Sigma$ is well-defined in all other
dimensions. The results show a discrepancy with respect to the prediction.
The discrepancy becomes smaller in higher dimensions. This suggests that the
study of flow invariants in non-unitary theories cannot be completely devoid
of meaning, but presumably the expression of the flow invariant is
not precise. The correct expression is found in section 7.

\begin{table}[tbp]
\begin{center}
\begin{tabular}{|c|c|c|}
\hline
$\gamma_+$ & $R_7$ & $R_9$ \\ \hline
1 & 1.41636 & 1.19275 \\ 
2 & 1.43223 & 1.18973 \\ 
3 & 1.44277 & 1.18614 \\ 
4 & 1.44711 & 1.18570 \\ 
6 & 1.44934 & 1.18561 \\ 
8 & 1.44977 & 1.18560 \\ 
10 & 1.44989 & 1.18560 \\ 
20 & 1.44997 & 1.18560 \\ 
50 & 1.44997 & 1.18560 \\ \hline
\end{tabular}
\end{center}
\caption{Ratio between calculated and predicted values of $\Sigma_n$ in
dependence of $m_1/m_2$ in 7 and 9 dimensions.}
\end{table}

In table 2 we report results in 7 and 9 dimensions, for various values of $%
r=\gamma_+^2$, to check if $\Sigma$ depends on the path connecting the UV
and IR fixed points. We see that the result does depend on $\gamma_+$, and
this dependence becomes smaller for high values of $\gamma_+$. This is
another confirmation that the proposed formula for $\Sigma$ cannot be
correct, but also that it must be somewhat close to the correct expression.

\section{Third test of the prediction}

\setcounter{equation}{0}

In the third test, we consider the fermionic higher-derivative theory with
lagrangian 
\[
{\cal L}=\bar\psi(\partial\!\!\!\slash+m_1) (\partial\!\!\!\slash%
+m_2)(\partial\!\!\!\slash+m_3)\psi. 
\]
The stress tensor can be computed with the method of section 5. This
problem, in four dimensions, was also considered in \cite{spin2}. The
formula (2.5) of \cite{spin2}, however, contains a mistake, since the term $%
\bar{\psi}\overleftarrow{\partial \!\!\!\slash}(\gamma _{\mu }%
\overleftrightarrow{\partial _{\nu }}+\gamma _{\nu }\overleftrightarrow{%
\partial _{\mu }})\partial \!\!\!\slash\psi$ was neglected. The expression
given there does not agree with the conformally-invariant coupling of the
theory to external gravity. We repeat the derivation here in arbitrary
dimensions.

The stress tensor is found by writing the most general linear combination of
terms, with free parameters as coefficients. The coefficients can be fixed
in three steps. In the first step, conservation is imposed up to terms
proportional to the field equations. In the second step, the condition of
vanishing trace is imposed up to terms proportional to the field equations.
After the first two steps two undetermined parameters survive. In the final
step, the expression of the stress tensor up to total derivatives is
considered, and matched with the result obtained by differentiating the
action with respect to the metric tensor. The non-minimal couplings can be
neglected in this procedure. We find, 
\[
T_{\mu\nu}=\left. e^a_{\{\mu}{\frac{\delta {\cal L}}{\delta e^a_{\nu\}}}}
\right|_{{\rm flat}}=\left. e^a_{\{\mu}{\frac{\delta }{\delta e^a_{\nu\}}}}%
\left(e\bar\psi D\!\!\!\!\slash^3\psi \right)\right|_{{\rm flat}}+{\rm t.}%
{\rm d.}= \bar\psi\left[2\partial_\mu\partial_\nu\partial\!\!\!\slash
+{\frac{1}{2}}(\gamma_\mu\partial_\nu+\gamma_\nu\partial_\mu)\Box\right]\psi
+{\rm t.}{\rm d.} 
\]
where ``t.d.'' means total derivatives. We have two terms and therefore two
new conditions. Consequently, the third step fixes the surviving parameters
and gives a unique answer.

Observe that, using this procedure it is not necessary to know the complete
expression of the conformally-invariant coupling to external gravity, i.e.
the analogue of formulas (\ref{uno}) and (\ref{due}). Very likely, the
result (\ref{glu}) is sufficient to determine the complete
conformally-invariant coupling. These considerations will not be pursued
here.

In the massless theory, the stress tensor reads 
\begin{eqnarray}
T_{\mu \nu } &=&{\frac{1}{4}}\left( \bar{\psi}(\gamma _{\mu }\partial _{\nu
}+\gamma _{\nu }\partial _{\mu })\Box \psi -\Box \bar{\psi}(\gamma _{\mu }%
\overleftarrow{\partial _{\nu }}+\gamma _{\nu }\overleftarrow{\partial _{\mu
}})\psi\right)-{\frac{1}{4}}\bar{\psi}\overleftarrow{\partial \!\!\!\slash}%
(\gamma _{\mu }\overleftrightarrow{\partial _{\nu }}+\gamma _{\nu }%
\overleftrightarrow{\partial _{\mu }})\partial \!\!\!\slash\psi  \nonumber \\
&+& {\frac{n+2}{4(n-2)}} \left(\Box \bar{\psi}(\gamma _{\mu }\partial _{\nu
}+\gamma _{\nu }\partial _{\mu })\psi- \bar{\psi}(\gamma _{\mu }%
\overleftarrow{\partial _{\nu }}+\gamma _{\nu }\overleftarrow{\partial _{\mu
}})\Box \psi\right) +{\frac{1}{(n-2)}} \partial _{\alpha }\bar{\psi}(\gamma
_{\mu }\overleftrightarrow{\partial _{\nu }}+\gamma _{\nu }%
\overleftrightarrow{\partial _{\mu }})\partial _{\alpha }\psi  \nonumber \\
&+&{\frac{1}{(n-1)(n-2)}} (\partial _{\mu }\bar{\psi}\overleftrightarrow{%
\partial \!\!\!\slash}\partial _{\nu }\psi +\partial _{\nu }\bar{\psi}%
\overleftrightarrow{\partial \!\!\!\slash}\partial _{\mu }\psi) -{\frac{n}{%
(n-1)(n-2)}} (\bar{\psi}\overleftarrow{\partial \!\!\!\slash}\partial _{\mu
}\partial _{\nu }\psi -\partial _{\mu }\partial _{\nu }\bar{\psi}\partial
\!\!\!\slash\psi )  \nonumber \\
&-&{\frac{1}{n-1}}(\bar{\psi}\partial \!\!\!\slash \partial _{\mu }\partial
_{\nu }\psi -\partial _{\mu }\partial _{\nu }\bar{\psi}\overleftarrow{%
\partial \!\!\!\slash}\psi ) +{\frac{n}{(n-1)(n-2)}}\delta _{\mu \nu } (\bar{%
\psi}\overleftarrow{\partial \!\!\!\slash}\Box \psi -\Box \bar{\psi}\partial
\!\!\!\slash\psi )  \nonumber \\
&-&{\frac{2}{(n-1)(n-2)}}\delta _{\mu \nu } \partial _{\alpha }\bar{\psi}%
\overleftrightarrow{\partial \!\!\!\slash}\partial _{\alpha }\psi+{\frac{1}{%
n-1}}\,\delta_{\mu\nu} (\bar\psi\Box\partial\!\!\!\slash\psi-\Box\bar\psi 
\overleftarrow{\partial\!\!\!\slash}\psi),  \label{glu}
\end{eqnarray}
The term proportional to the free-field equations can be fixed by imposing
conservation in the massive case (see below).

The mass-independent part of the stress tensor contributes to the trace with 
\[
\Theta={\frac{3}{2}}(\bar\psi\Box \partial\!\!\!\slash\psi-\Box\bar\psi
\partial\!\!\!\slash\psi). 
\]
The central charge $c_n$ is found from the two-point function (\ref{twop}).
The result is 
\[
c_n=-2^{[n/2]-1}\,{\frac{n^2+n-18}{n-2}} 
\]
The prediction (\ref{predic1}) reads in this case 
\begin{equation}
\Sigma_n=-{\frac{2^{[n/2]-1}(n^2+n-18)\,\Gamma(n/2+1) }{\pi^{n/2}(n+1)(n-2)}}%
.  \label{Pn}
\end{equation}

In the massive case, the other contributions to the stress tensor can be
written using the formulas for the theories $\bar\psi\Box \psi$ and $%
\bar\psi\partial\!\!\!\slash\psi$. We have 
\begin{eqnarray*}
\Delta T_{\mu\nu}&=&{\frac{1}{4}}(m_1+m_2+m_3) \left(\bar\psi\gamma_\mu 
\overleftrightarrow{\partial_\nu}\partial\!\!\!\slash\psi+
\bar\psi\gamma_\nu \overleftrightarrow{\partial_\mu}\partial\!\!\!\slash%
\psi- \bar\psi\overleftarrow{\partial\!\!\!\slash}\gamma_\mu 
\overleftrightarrow{\partial_\nu}\psi- \bar\psi\overleftarrow{\partial\!\!\!%
\slash}\gamma_\nu \overleftrightarrow{\partial_\mu}\psi \right)  \nonumber \\
&+&{\frac{1}{4}}(m_1m_2+m_1m_3+m_2m_3)\left(\bar\psi\gamma_\mu 
\overleftrightarrow{\partial_\nu}\psi+ \bar\psi\gamma_\nu 
\overleftrightarrow{\partial_\mu}\psi \right)
\end{eqnarray*}
plus a couple of terms proportional to the field equations, which we can
omit. In the end, we find the following trace: 
\begin{eqnarray*}
\Theta&=&-(m_1+m_2+m_3)(\bar\psi\Box\psi+ \bar\psi\overleftarrow{%
\partial\!\!\!\slash}\partial\!\!\!\slash\psi+ \Box\bar\psi\psi)  \nonumber
\\
&-& (m_1m_2+m_1m_3+m_2m_3)(\bar\psi\partial\!\!\!\slash\psi-\bar\psi 
\overleftarrow{\partial\!\!\!\slash}\psi)-3m_1m_2m_3 \bar\psi\psi.
\end{eqnarray*}
The improvement operator is ${\cal O}=\Box(\bar\psi\psi)$.

\begin{table}[tbp]
\begin{center}
\begin{tabular}{|cc|cc|cc|}
\hline
$n$ & $R$ & $n$ & $R$ & $n$ & $R$ \\ \hline
~4 & sing. ~ & ~ 10 & 1.14773~ & ~ 70 & 1.00308~ \\ 
~5 & 2.36284 ~ & ~ 12 & 1.09900~ & ~ 100 & 1.00153~ \\ 
~6 & 1.60279 ~ & ~ 15 & 1.06230~ & ~ 150 & 1.00069~ \\ 
~7 & 1.36553 ~ & ~ 20 & 1.03511~ & ~ 200 & 1.00039~ \\ 
~8 & 1.25320 ~ & ~ 30 & 1.01593~ & ~ 300 & 1.00017~ \\ 
~9 & 1.18887 ~ & ~ 45 & 1.00726~ & ~ 500 & 1.00006~ \\ \hline
\end{tabular}
\end{center}
\caption{True versus predicted value in various dimensions.}
\end{table}

We consider the case $m_1=m_2=-m_3\equiv m$, for simplicity, where the
porpagator is 
\[
\langle\bar\psi(x)\,\psi(0)\rangle= mG_{n,1}(x)-{\frac{x\!\!\!\slash}{|x|}}%
G^{\prime}_{n,1}(x) 
\]
and $G_{n,1}(x)$ is the same as in (\ref{gn1}).

The results are reported table 3 and are very similar to those of the scalar
theory studied in section 5. Again, there is a discrepancy with respect to
the prediction. The discrepancy, however, decreases when the space-time
dimension increases. In dimension 500, we have a 0.006\% discrepancy.

\section{The solution of the puzzle}

\setcounter{equation}{0}

The numerical results presented in the previous sections do not reproduce
the predictions exactly. Sometimes, such as for $n=4,5$ in the scalar theory
of section 5 and for $n=4$ in the fermion theory of section 6, the flow
integral is ill-defined. This happens because the correlators $\langle
\Box\varphi^2\,\Box\varphi^2\rangle $ or $\langle
\Box\bar\psi\psi\,\Box\bar\psi\psi\rangle $ vanish in one or more points. We
know that, on the other hand, in physical, unitary quantum field theories,
these correlators are positive.

We have already discussed that there are good reasons to believe that the
flow invariant is meaningful in the mathematically well-defined theories.
The very fact that the results of the previous sections are in most cases
close to the predictions, althought not equal to those, supports this
consideration. The point is that we have not used the correct formula for
the flow invariant. The solution is as follows.

We have required that $\sigma(x)$ be invariant under the symmetry (\ref{inva}%
), but this requirement is too strong. Indeed, we just need that the
integral of $\sigma(x)$ be invariant under this symmetry. The correct
formula reads 
\begin{equation}
\sigma_n=\int {\rm d}^{n}x \,|x|^n \langle \Theta(x) \, \Theta(0)\rangle
-M^t N^{-1}M,  \label{sigma2}
\end{equation}
where $M_i$ and $N_{ij}$ are the vector and matrix defined by 
\[
M_i=\int {\rm d}^{n}x \,|x|^n \langle \Theta(x) \, {\cal O}_i(0)\rangle,
~~~~~~~~~~~N_{ij}=\int {\rm d}^{n}x \,|x|^n \langle {\cal O}_i(x) \, {\cal O}%
_j(0)\rangle 
\]
The invariant $\sigma_n$ is more general than $\Sigma_n$, since it satisfies
nothing more that the minimum symmetry requirements. The invariance of (\ref
{sigma2}) under (\ref{inva}) is easy to prove and I leave this as an
exercise for the reader.

We have produced two expressions, $\sigma_n$ and $\Sigma_n$, which are both
invariant under (\ref{inva}). This means that the symmetry (\ref{inva}) does
not fix the invariant uniquely and that we need a more powerful principle.
The answer is a minimum principle stating that 
\begin{equation}
\sigma_n=\min_a\int {\rm d}^n x \, |x|^n\, \langle \Theta_a (x) \,\Theta_a
(0) \rangle,  \label{porta}
\end{equation}
where 
\[
\Theta_a=\Theta+a_i{\cal O}_i, 
\]
and $a_i$ are arbitrary constants. Denoting by $\bar a_i$ the constants
minimizing the expression above, we find the solution 
\[
\bar a_i=-(N^{-1})_{ij}M_j, 
\]
whence the result (\ref{sigma2}) follows.

Formula (\ref{porta}) means that we have to minimize the integral of $%
\langle\Theta\,\Theta\rangle$ in the entire space of improved $\Theta$'s. In
unitary theories, there always exist a non-negative minimum, since %
\hbox{$\langle \Theta_a (x) \,\Theta_a (0) \rangle\ge 0$}. The minimum is
zero if and only if the theory is conformal.

Indeed, in a unitary theory, the condition of criticality in the presence of
improvement terms is not defined by $\Theta=0$, which is meaningless, but by
the equality $\sigma_n=0$, which is equivalent to say that there exist
constants $\bar a_i$ such that 
\[
\Theta=-\bar a_i\, {\cal O}_i. 
\]
The proof of this fact is straightforward. Putting $\sigma_n=0$ in
expression (\ref{porta}) and using the fact that $\langle
\Theta_a\,\Theta_a\rangle\geq 0$, we see that there exist constants $\bar
a_i $ such that $\langle\Theta_{\bar a}(x)\, \Theta_{\bar a}(0)\rangle\equiv
0$. In a unitary theory this means that the operator $\Theta_{\bar a}$
vanishes, which implies the statement.

We can prove that in unitary theories $\sigma_n\geq \Sigma_n$. For
simplicity, we consider the case of a single improvement operator ${\cal O}$%
. The proof is a standard application of the Schwarz-H\"older inequality. We
have, for arbitrary functions $f$ and $g$ and an arbitrary constant $a$, 
\begin{equation}
0\leq \int (f+a g)^2\,{\rm d}\mu = \int f^2 \, {\rm d}\mu + 2 a\int fg\,{\rm %
d}\mu + a^2 \int g^2\,{\rm d}\mu,  \label{shw}
\end{equation}
where ${\rm d}\mu$ denotes the integration measure. Since (\ref{shw}) holds
for arbitrary $a$, the discriminant must be negative, or zero: 
\[
\left(\int fg\,{\rm d}\mu\right)^2 \leq \left(\int f^2 \, {\rm d}%
\mu\right)\left(\int g^2\,{\rm d}\mu\right). 
\]
In unitary theories, $\langle{\cal O}(x)\,{\cal O}(0)\rangle>0$ and choosing 
\[
g(x)=\sqrt{\langle{\cal O}(x)\,{\cal O}(0)\rangle}, ~~~~~~ f(x)={\frac{1}{%
g(x)}}\langle\Theta(x)\,{\cal O}(0)\rangle, 
\]
and ${\rm d}\mu={\rm d}^nx\, |x|^n$, we have 
\[
{\frac{\left( \int {\rm d}^nx\, |x|^n\, \langle\Theta(x)\,{\cal O}%
(0)\rangle\right)^2 }{\int {\rm d}^nx\, |x|^n\, \langle{\cal O}(x)\,{\cal O}%
(0)\rangle}} \leq \int {\rm d}^nx\, |x|^n\, {\frac{\left(\langle\Theta(x)\,%
{\cal O}(0)\rangle\right)^2 }{\langle{\cal O}(x)\,{\cal O}(0)\rangle}}. 
\]
Adding $-\int {\rm d}^nx\, |x|^n\, \langle\Theta(x)\,\Theta(0)\rangle$ to
both sides, we get the desired result, $\sigma_n\geq \Sigma_n$.

This conclusion is quite reasonable. The integrand $\sigma(x)$ of $\Sigma_n$
is the minimum value of the correlator $\langle\Theta_a(x)\,\Theta_a(0)%
\rangle$. However, the value $\bar a$ at which the correlator $%
\langle\Theta_a(x)\,\Theta_a(0)\rangle$ (not its integral) is minimum, is
point-dependent, $\bar a=\bar a(x)$. Therefore $\Theta_{\bar a}$ is
non-local. If we allow $a$ to be point-dependent, we are minimizing over a
much larger space and therefore we get a smaller value $\Sigma_n\leq
\sigma_n $. The mentioned non-locality is the ultimate reason why $\Sigma_n$
cannot be the correct expression for the flow invariant.

In non-unitary theories, $\langle{\cal O}(x)\,{\cal O}(0)\rangle$ does not
have, in general, a definite sign. If, however, it happens that it is
identically negative, then we can repeat the above argument by replacing the
correlator $\langle{\cal O}(x)\,{\cal O}(0)\rangle$ with its absolute value,
in the definition of the function $g$. It is easy to verify that we get the
inversed inequality $\sigma_n\leq \Sigma_n$. In the bosonic model considered
in section 5, the correlator $\langle{\cal O}(x)\,{\cal O}(0)\rangle$ is
identically positive. This means that the numerical value of $\Sigma_n$
should be smaller than the predicted value $\sigma_n$. Since both are
negative, this implies $R\geq 1$. This is always verified: see tables 1 and
2. In the fermionic model of section 6 the situation is similar, when $%
m_1=m_2=-m_3$, as we see from table 3.

\medskip

In conclusion, the prediction (\ref{predic1}) should be replaced by 
\begin{equation}
\sigma_n=\Delta c_n\, {\frac{\Gamma(n/2+1)}{\pi^{n/2}\, (n+1)}}.
\label{predic2}
\end{equation}
Observe that the new formula (\ref{sigma2}) is easier to compute, since each
separate integral can be reduced to elementary integrals of rational
functions in momentum space.

\medskip

We now reconsider the bosonic model of section 5 and test the new prediction
in the case $m_1=m_2=m$.

We have 
\[
\sigma _{n}=\alpha _{n}-{\frac{\beta _{n}^{2}}{\gamma _{n}}}, 
\]
where 
\begin{eqnarray*}
\alpha _{n} &=&\int {\rm d}^{n}x\,|x|^{n}\,\langle \Theta (x)\,\Theta
(0)\rangle =\int {\frac{{\rm d}^{n}p}{(2\pi )^{n}}}(-1)^{n/2}{\frac{%
(4m^{4}p^{4}+16m^{4}p^{2}+8m^{4})I_{n}(p)+4m^{4}p^{2}J_{n}(p)}{%
(p^{2}+m^{2})^{2}}}, \\
\beta _{n} &=&2n(n-1)\int {\rm d}^{n}x\,|x|^{n-2}\,\langle \Theta
(x)\,\varphi ^{2}(0)\rangle =-8m^{4}n(n-1)\int {\frac{{\rm d}^{n}p}{(2\pi
)^{n}}}(-1)^{n/2-1}{\frac{I_{n}^{(2)}(p)}{(p^{2}+m^{2})}}, \\
\gamma _{n} &=&4n(n-1)(n-2)^{2}\int {\rm d}^{n}x\,|x|^{n-4}\,\langle \varphi
^{2}(x)\,\varphi ^{2}(0)\rangle =8n(n-1)(n-2)^{2}\int {\frac{{\rm d}^{n}p}{%
(2\pi )^{n}}}{\frac{(-1)^{n/2}I_{n}^{(1)}(p)}{(p^{2}+m^{2})^{2}}},
\end{eqnarray*}
and 
\begin{eqnarray*}
I_{n}(p) &\equiv &\left( {\frac{\partial ^{2}}{\partial p^{2}}}\right) ^{n/2}%
{\frac{1}{(p^{2}+m^{2})^{2}}},~~~~~~~~~J_{n}(p)\equiv \left( {\frac{\partial
^{2}}{\partial p^{2}}}\right) ^{n/2}{\frac{p^{2}}{(p^{2}+m^{2})^{2}}}, \\
I_{n}^{(1)}(p) &\equiv &\left( {\frac{\partial ^{2}}{\partial p^{2}}}\right)
^{n/2-2}{\frac{1}{(p^{2}+m^{2})^{2}}},~~~~~~~~~I_{n}^{(2)}(p)\equiv \left( {%
\frac{\partial ^{2}}{\partial p^{2}}}\right) ^{n/2-1}{\frac{1}{%
(p^{2}+m^{2})^{2}}}.
\end{eqnarray*}
These expressions can be worked out with a certain algebraic effort: 
\begin{eqnarray*}
I_{n}(p) &=&2^{n-1}n!(-1)^{n/2}(m^{2})^{n/2-2}{\frac{%
p^{4}(n/2-1)-2m^{2}p^{2}(n/2+1)+m^{4}(n/2+1)}{(p^{2}+m^{2})^{n+2}}}, \\
J_{n}(p) &=&2^{n}n!(-1)^{n/2+1}(m^{2})^{n/2-1}{\frac{%
p^{4}(n/4)-m^{2}p^{2}(n/2+1)+m^{4}(n/4)}{(p^{2}+m^{2})^{n+2}}}, \\
I_{n}^{(1)}(p) &=&2^{n-4}(n-3)!(-1)^{n/2}(m^{2})^{n/2-2}{\frac{1}{%
(p^{2}+m^{2})^{n-2}}}, \\
I_{n}^{(2)}(p) &=&2^{n-2}(n-2)!(-1)^{n/2}(m^{2})^{n/2-2}{\frac{%
p^{2}(n/2-1)-m^{2}(n/2)}{(p^{2}+m^{2})^{n}}}.
\end{eqnarray*}

The integrals are elementary and give 
\[
\alpha_n={\frac{2\Gamma(n/2+1)}{(n+1)\pi^{n/2}}},~~~~ \beta_n=-{\frac{%
2\Gamma(n/2+1)}{m^2\pi^{n/2}}},~~~~ \gamma_n={\frac{(n-2)\Gamma(n/2+1)}{%
m^4\pi^{n/2}}} 
\]
Finally, we find 
\[
\sigma_n=-{\frac{2(n+4)\Gamma(n/2+1)}{(n+1)(n-2)\pi^{n/2}}}, 
\]
in agreement with the prediction.

We now continue the analysis of the scalar model of section 5, but set $%
m_1\neq m_2$. We find an unexpected behavior. It would be natural to expect
that the value of the flow integral does not depend of the ratio $m_1/m_2$,
i.e. on the particular trajectory connecting the same UV and IR fixed points
(see point $iii$ of sect. 1.2). It turns out, however, that this is not the
case. The value of the flow integral does know about the trajectory. It is
minimal and equal to (\ref{predic2}) on a priviledged trajectory, which is,
in our case, precisely the trajectory with $m_1=m_2$.

I have computed $\sigma_n$ numerically for various values of $m_1/m_2$ in 7
and 9 dimensions. The results, normalized to the predicted value appearing
on the right-hand side of (\ref{predic2}), are shown in table 4.

\begin{table}[tbp]
\begin{center}
\begin{tabular}{|c|c|c|}
\hline
$\gamma_+$ & $R_7$ & $R_9$ \\ \hline
1 & 1.000000 & 1.000000 \\ 
2 & 0.924873 & 0.906629 \\ 
3 & 0.778893 & 0.742547 \\ 
4 & 0.681417 & 0.640275 \\ 
6 & 0.580400 & 0.537945 \\ 
8 & 0.530913 & 0.488573 \\ 
10 & 0.501515 & 0.459344 \\ 
20 & 0.440888 & 0.399087 \\ 
50 & 0.397027 & 0.355388 \\ \hline
\end{tabular}
\end{center}
\caption{Ratio between calculated and predicted values of $\sigma_n$ in
dependence of $m_1/m_2$ in 7 and 9 dimensions.}
\end{table}

We see that the ratio $R$ is not constant, but decreases when the $m_1/m_2$
departs from 1 (the result is clearly invariant under $m_1/m_2\rightarrow
m_2/m_1$). Recalling that the value of the integral is negative, we conclude
that the flow integral is minimal, and equal to the prediction, for $m_1=m_2$%
.

As a further check, we can compute the flow integral $\sigma$ exactly in $p$%
-space for $n=4$, as a function of $m_1^2/m_2^2=r^2$. The result reads 
\[
\sigma(r)=-6\, {\frac{(r^2-1)^2(3r^4- 26r^2 +3 ) + ( r^8 + 18r^6-
18r^2-1)\ln r^2 - 10r^2(r^4+1)\ln^2 r^2}{5 \pi^2 (r^2-1)^3 (r^2 \ln r^2+ \ln
r^2- 2r^2 +2 )}} 
\]
and is plotted in figure 2, with $x=\ln r^2$ in the abscissa. The minimum is
at $r=1$, where it equals the expected value $\sigma_4=-16/(5\pi^2)$. The
maximum value, for $r=0$ and $r=\infty$, is $-6/(5\pi^2)$.

\begin{figure}[tbp]
\begin{center}
\let\picnaturalsize=N \ifx\nopictures Y\else{\ifx\epsfloaded Y\else\fi
\global\let\epsfloaded=Y \centerline{\ifx\picnaturalsize N\epsfxsize
4in\fi \epsfbox{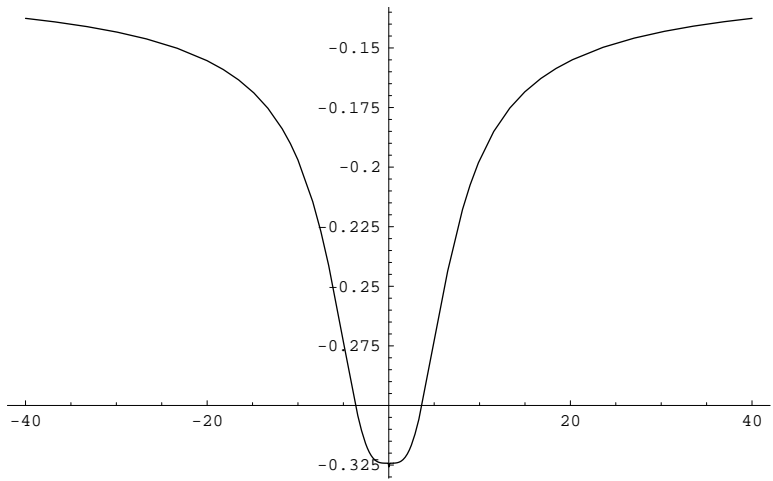}}}\fi
\end{center}
\caption{Plot of $\sigma_n$ for the bosonic model of sect. 5 in $n=4$.}
\end{figure}

An exact calculation can also be done in the fermion model of section 6, in
the case $m_1=m_2=-m_3=m$. Since, however, it is rather lengthy, I have
preferred to check the prediction numerically. In all cases the agreement $%
R=1$ has been found, up to the sixth decimal figure. The results are given
in table 5, normalized to the prediction $P_n$, that is to say the
right-hand side of (\ref{Pn}).

\begin{table}[tbp]
\begin{center}
\begin{tabular}{||c|c|c|c|c||c|c|c|c|c||}
\hline
$n$ & $-\alpha_n/P_n$ & $-\beta_n/P_n$ & $-\gamma_n/P_n$ & $\sigma_n/P_n$ & $%
n$ & $-\alpha_n/P_n$ & $-\beta_n/P_n$ & $-\gamma_n/P_n$ & $\sigma_n/P_n$ \\ 
\hline
4 & 9 & 10 & 10 & 1 & 20 & 2.55224 & 31.9701 & 287.731 & 1 \\ 
5 & 3 & 6 & 9 & 1 & 30 & 2.67105 & 51.3947 & 719.526 & 1 \\ 
6 & 5/2 & 7 & 14 & 1 & 45 & 2.76608 & 80.9708 & 1740.87 & 1 \\ 
7 & 2.36842 & 8.42105 & 21.0526 & 1 & 70 & 2.84249 & 130.645 & 4441.92 & 1
\\ 
8 & 7/3 & 10 & 30 & 1 & 100 & 2.88693 & 190.459 & 9332.51 & 1 \\ 
9 & 7/3 & 35/3 & 245/6 & 1 & 150 & 2.92312 & 290.311 & 21483. & 1 \\ 
10 & 2.34783 & 13.3913 & 53.5652 & 1 & 200 & 2.94176 & 390.235 & 38633.2 & 1
\\ 
12 & 2.3913 & 16.9565 & 84.7826 & 1 & 300 & 2.96079 & 590.158 & 87933.5 & 1
\\ 
15 & 2.45946 & 22.4865 & 146.162 & 1 & 500 & 2.97629 & 990.095 & 246534. & 1
\\ \hline
\end{tabular}
\end{center}
\caption{Results for the fermionic model of section 6 with $m_1=m_2=-m_3$.}
\end{table}

The situation is more complicated, when we consider other trajectories than $%
m_1=m_2=-m_3$. Taking $m_1=m_2$, the integral of (\ref{sigma2}) is a
function of $m_3$, plotted in figure 3. We see that the point $m_3=-1$ is an
extremum, but not a minimum of $\sigma_4$ (a maximum, in this case) and that
a minimum does not exist. There is, instead, a singularity for $%
m_3=-0.683155 $ and there exist other extrema beyond the singular point. One
of these is $m_1=m_2=m_3$, for which the ratio $\sigma_4/P_4$ equals $-135$,
a value which does not have a clear interpretation. The singularity visible
in figure 3 is due to a zero of $\gamma_4$ and is a clear sign that certain
non-unitary theories can be sufficiently bad to give unexpected problems. It
might be wise to restrict the set of non-unitary theories to the purely
bosonic ones, where there exists a notion of positive-definiteness for the
action. Eventually, we can include the supersymmetric non-unitary theories
having a positive-definite bosonic action. We naturally expect that the
boson-fermion pairing imposed by supersymmetry forces the fermionic sector
of the theory to be also well-behaved.

I have checked that $m_1=m_2=-m_3$ and $m_1=m_2=m_3$ are extrema in the full
space of the parameters $m_1$, $m_2$ and $m_3$. I have also extended the check
to various dimensions other than 4. I report here only a few results,
obtained numerically. In particular, at the point $m_1=m_2=-m_3=1$, we have 
\begin{eqnarray*}
{\frac{\partial(\alpha_4/P_4)}{\partial m_{1,2}}}&=&-20,~~~~~ {\frac{%
\partial(\beta_4/P_4)}{\partial m_{1,2}}}=0,~~~~~{\frac{\partial(%
\gamma_4/P_4)}{\partial m_{1,2}}}=20, \\
&&~~~~~~~~~~~~~~~~~~~~~~~~{\frac{\partial(\sigma_4/P_4)}{\partial m_{1,2}}}={%
\frac{\partial(\alpha_4/P_4)}{\partial m_{1,2}}}-2{\frac{\beta_4}{\gamma_4}}{%
\frac{\partial(\beta_4/P_4)}{\partial m_{1,2}}}+{\frac{\beta_4^2}{\gamma_4^2}%
} {\frac{\partial(\gamma_4/P_4)}{\partial m_{1,2}}}=0, \\
{\frac{\partial(\alpha_4/P_4)}{\partial m_{3}}}&=&-40,~~~~~ {\frac{%
\partial(\beta_4/P_4)}{\partial m_{3}}}=-10,~~~~~{\frac{\partial(%
\gamma_4/P_4)}{\partial m_{3}}}=20, ~~~~~{\frac{\partial(\sigma_4/P_4)}{%
\partial m_{3}}}=0,
\end{eqnarray*}
in dimension 4 and 
\begin{eqnarray*}
{\frac{\partial(\alpha_6/P_6)}{\partial m_{1,2}}}&=&-{\frac{14}{3}},~~~~~ {%
\frac{\partial(\beta_6/P_6)}{\partial m_{1,2}}}=0,~~~~~{\frac{%
\partial(\gamma_6/P_6)}{\partial m_{1,2}}}={\frac{56}{3}},~~~~~ {\frac{%
\partial(\sigma_6/P_6)}{\partial m_{1,2}}}=0, \\
{\frac{\partial(\alpha_6/P_6)}{\partial m_{3}}}&=&-{\frac{28}{3}},~~~~~ {%
\frac{\partial(\beta_6/P_6)}{\partial m_{3}}}=-7,~~~~~{\frac{%
\partial(\gamma_6/P_6)}{\partial m_{3}}}={\frac{28}{3}}, ~~~~~{\frac{%
\partial(\sigma_6/P_6)}{\partial m_{3}}}=0,
\end{eqnarray*}
in dimension 6.

\begin{figure}[tbp]
\begin{center}
\let\picnaturalsize=N \ifx\nopictures Y\else{\ifx\epsfloaded Y\else\fi
\global\let\epsfloaded=Y \centerline{\ifx\picnaturalsize N\epsfxsize
5in\fi \epsfbox{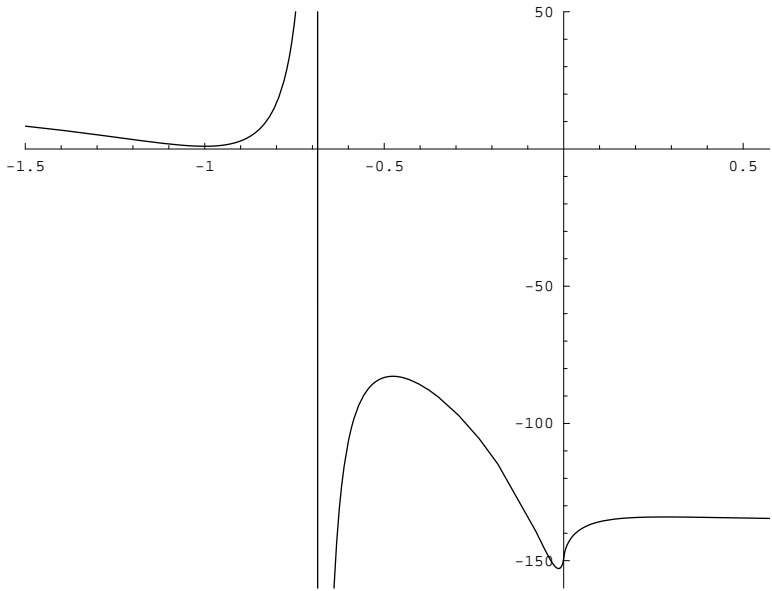}}}\fi
\end{center}
\caption{Plot of $\sigma_4/P_4$ versus $m_3$ for the fermionic model of
sect. 6 in $n=4$, with $m_1=m_2$.}
\end{figure}

\section{Conclusions}

\setcounter{equation}{0}

Collecting the information obtained so far, the final formula of the flow
invariant reads in unitary theories 
\begin{equation}
\sigma _{n}=\min_{m(\mu )}\,\min_{a}\int {\rm d}^{n}x\,|x|^{n}\,\langle
\Theta _{a}(x)\,\Theta _{a}(0)\rangle ,  \label{final}
\end{equation}
where the minimization is performed both in the space of improved stress
tensors and in the space of trajectories $m(\mu )$ relating the dimensioned
parameters of the theory. The invariant is universal and proportional to $%
\Delta a$ in classically-conformal theories (marginally-relevant flows),
where there is no need to minimize over the set of trajectories $m(\mu )$,
because there is only the dynamical scale $\mu $. It is proportional to $%
\Delta c$ in flows of strictly 
relevant operators, where $\mu $ is absent. There is
evidence to claim that the flow invariant is universal in the class of flows
with $\Delta a=\Delta c$. When $\Delta a\neq \Delta c$ the formula is still
expected to give a characterization of the flow, in the sense that the minimum
principle selects priviledged flows among the flows connecting the
same fixed points. These flows are special, because in the examples
considered in this paper the integral appearing in $\sigma_n$, 
calculated along the priviledged trajectory, equals 
the predicted value $\Delta c$. 

Several problems remain open. The relation between the
flow invariant and the central charges $a$ and $c$ has to be clarified in
the general flows (see point $ii$ of sect. 1.2). A\ possible way to shed
light on the open questions might be to reconsider the issues studied here
from the point of view of 
the wilsonian exact renormalization-group approach. Here I have studied
mostly gaussian theories, but the results are expected to be more general,
since they include all the cases treated so far \cite{noi,noi2,athm,cea}.

In non-unitary theories with a positive-definite action, we have found that
the minimum still exists. Nevertheless, the condition of
positive-definiteness of the action is meaningful only for bosonic theories
and it is not straightforward to define the acceptable non-unitary fermionic
theories. We have found that in such theories the minimum might not exist,
and the prediction holds at an extremum of the flow integral of (\ref{final}
). Singularities might limit the acceptable region. Typically, other extrema
exist outside of it, but the value of the integral (\ref{final}) in those
points does not seem to have a clear interpretation. The privileged
trajectory is $m_1=m_2$ in the bosonic model of section 5, and $m_1=m_2=-m_3$
in the fermionic model of section 6.

Our initial expectations have been updated in two relevant points. First, it
is now clear how to define the flow invariant in the presence of improvement
terms for the stress tensor. A natural minimum principle selects the correct
formula. Second, the flow invariant is not independent of the trajectory in
the space of dimensioned parameters. This might also suggest that the name
``flow invariant'', conceived on the basis of the initial expectations (that
the integral (\ref{sigma2}) was completely independent on the trajectory) is
not completely justified. Equivalently, we can say that flow invariance is
trivial, since the minimum $\sigma _{n}$ of (\ref{final}) is obviously
independent of the trajectory, given that it is obtained by minimizing (\ref
{sigma2}) over all trajectories. 
I expect that the form (\ref{final}) of the flow invariant
applies also to the Standard Model, in which case we might be able to relate
the Higgs mass to $\Lambda _{{\rm QCD}}$. Finally, it can be speculated that
a more sophisticated invariant might be written, by giving a suitable weight
to each trajectory and functionally integrating over all trajectories,
instead of minimising.

\medskip {\bf Acknowledgements} \vskip .2truecm

I am grateful to P. Menotti and R. Rattazzi for useful discussions.

\end{document}